\documentclass[twocolumn]{article}

\usepackage[margin=1in]{geometry}
\usepackage{amsmath}
\usepackage{amssymb}
\usepackage{xcolor}
\usepackage{cite}
\usepackage{hyperref}

\usepackage{picinpar,moresize,xfrac,graphpap,dcolumn,wrapfig,graphicx}

\usepackage{subcaption}
\usepackage{stackengine}

\usepackage{tikz,pgfplots}
\usepackage{tikz-cd}
\usepackage{pgf}
\usepgfplotslibrary{colormaps}
\usetikzlibrary{pgfplots.colormaps}
\usetikzlibrary{arrows,automata,positioning,backgrounds}
\usepackage{rotating}
\pgfplotsset{compat=newest}
\pgfplotsset{plot coordinates/math parser=false}
\newlength\figureheight
\newlength\figurewidth

\usepackage{soul,array,calc,url,ragged2e,graphpap}
\usepackage{booktabs} 
\usepackage{tabularx}
\usepackage{colortbl}
\usepackage{multirow}
\usepackage{hyphenat}
\usepackage{transparent}
\usepackage{enumerate}
\usepackage{mathrsfs}

\usepackage[figure]{hypcap}
\usepackage{mathtools}
\mathtoolsset{centercolon}
\usepackage{nicefrac}
\usepackage{units}
\usepackage[normalem]{ulem}
\usepackage{cancel}
\usepackage{pbox}
\usepackage{multicol}
\usepackage{cases}
\usepackage[all]{xy}
\usepackage{nicematrix}

\usepackage{cleveref}
\crefmultiformat{subequation}%
{\edef\crefstripprefixinfo{#1}(#2#1#3}%
{,#2\crefstripprefix{\crefstripprefixinfo}{#1}#3)}%
{,#2\crefstripprefix{\crefstripprefixinfo}{#1}#3}%
{,#2\crefstripprefix{\crefstripprefixinfo}{#1}#3)}

\makeatletter
\DeclareFontFamily{OMX}{MnSymbolE}{}
\DeclareSymbolFont{MnLargeSymbols}{OMX}{MnSymbolE}{m}{n}
\SetSymbolFont{MnLargeSymbols}{bold}{OMX}{MnSymbolE}{b}{n}
\DeclareFontShape{OMX}{MnSymbolE}{m}{n}{
    <-6>  MnSymbolE5
   <6-7>  MnSymbolE6
   <7-8>  MnSymbolE7
   <8-9>  MnSymbolE8
   <9-10> MnSymbolE9
  <10-12> MnSymbolE10
  <12->   MnSymbolE12
}{}
\DeclareFontShape{OMX}{MnSymbolE}{b}{n}{
    <-6>  MnSymbolE-Bold5
   <6-7>  MnSymbolE-Bold6
   <7-8>  MnSymbolE-Bold7
   <8-9>  MnSymbolE-Bold8
   <9-10> MnSymbolE-Bold9
  <10-12> MnSymbolE-Bold10
  <12->   MnSymbolE-Bold12
}{}
\let\llangle\@undefined
\let\rrangle\@undefined
\DeclareMathDelimiter{\llangle}{\mathopen}%
                     {MnLargeSymbols}{'164}{MnLargeSymbols}{'164}
\DeclareMathDelimiter{\rrangle}{\mathclose}%
                     {MnLargeSymbols}{'171}{MnLargeSymbols}{'171}
\makeatother

\newcommand*{\vertbar}{\rule[-1ex]{0.5pt}{2.5ex}}
\newcommand*{\horzbar}{\rule[.5ex]{2.5ex}{0.5pt}}

\usepackage{fancyvrb,listings}
\usepackage{algorithm}
\usepackage{algorithmicx}
\usepackage{algpseudocode}

\algrenewcommand\alglinenumber[1]{\footnotesize #1:}
\makeatletter  
 \renewcommand{\ALG@name}{\small Algorithm} 
\makeatother 

\newtheorem{definition}{Definition}
\newtheorem{problem}{Problem}

\newcommand{\probref}[1]{\textup{Problem~\ref{#1}}}

\def\etal{\emph{et al.}}

\def\II{\mathbb{I}}

\def\RR{\mathbb{R}}
\def\SS{\mathbb{S}}

\def\bK{\mathbf{K}}

\def\bS{\mathbf{S}}

\def\cE{\mathcal{E}}

\def\cP{\mathcal{P}}

\usepackage{bbm}
\DeclareSymbolFont{bbold}{U}{bbold}{m}{n}
\DeclareSymbolFontAlphabet{\mathbbold}{bbold}

\def\det{\operatorname{det}}

\def\Mhat{\widehat{M}}

\title{Adaptive Surface Meshes from Harmonic Maps}

\author{Nicolas Nebel \and Albert Chern}
\begin{document}

\twocolumn[{
\renewcommand\twocolumn[1][]{#1}%
\maketitle
\includegraphics[width=\textwidth]{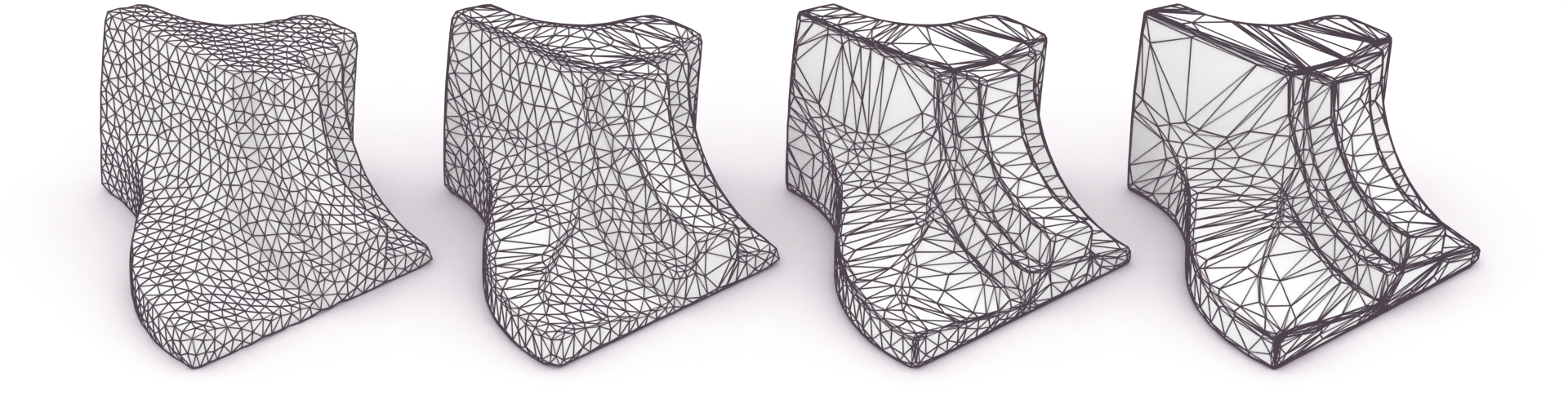}
\captionof{figure}{The harmonic map onto a surface with a custom Riemannian structure produces a mesh that efficiently and adaptively approximates a given shape (right). This shape-approximating mesh is found by providing an initial mesh (left) and continuously flowing it (left to right) under a heat diffusion-like flow.}\label{fig:teaser}\par\bigskip}]

\begin{abstract}
   We present a novel shape-approximating anisotropic re-meshing algorithm as a geometric generalization of the adaptive moving mesh method. Conventional moving mesh methods reduce the interpolation error of a mesh that discretizes a given function over a planar domain. Our algorithm, in contrast, optimizes the mesh's approximation of a curved surface; surfaces can be represented in various formats, such as a signed distance field. The optimization is achieved by continuously flowing the mesh without altering its topology, making the implementation simpler compared to other adaptive surface meshing techniques. The resulting optimal mesh can be interpreted as a harmonic map with respect to a metric using the shape operator. Furthermore, our approach can be tailored to target height fields by utilizing isotropic geometry.
\end{abstract}

\section{Introduction}

Finding an accurate, yet concise triangulation of an arbitrary surface is an important task in many areas of computer graphics and computer-aided design. 
Often times, one already has a topologically faithful mesh that otherwise insufficiently approximates an arbitrary shape (given by, for example, a signed distance field or height map).
We introduce a new method that iteratively evolves these existing meshes to approach ideal surface approximations.
Unlike previous methods, our iterative procedure does not require sophisticated operations that change the topology of a mesh.

In the context of optimizing planar meshes for function interpolation, this approach of evolving existing meshes is known as an \emph{adaptive moving mesh method} \cite{huang11}.
Our method generalizes moving mesh methods to curved surface approximation, beyond just functions on planar domains.

To achieve this geometric generalization, we re-visit the continuous theory for the moving mesh methods. Traditionally, the target surface is endowed with a \emph{monitor function} which determines the desired anisotropy of the mesh. In our re-formulation, the monitor function is defined as the positivized shape operator --- analogous to the positivized Hessian in a planar moving mesh method. The mesh flow is then described as an evolving map that maps from a manifold representing the topological mesh to the surface of interest. Using the monitor function as the Riemannian structure, we minimize the Dirichlet energy of the map, yielding a so-called \emph{harmonic map} that faithfully displays the desired anisotropy.
Intuitively, this moving mesh flow is merely the heat diffusion process associated with this Dirichlet energy.
Specifically, this paper makes the following contributions:
\begin{itemize}
    \item We generalize moving mesh theory to curved domains.
    \item We apply our generalized moving mesh method to the problem of surface approximation. Our implementation only requires an SDF and an initial mesh.
    \item We describe details in the implementation, including anisotropy control, boundary treatment, and per-face shape operator calculation.
\end{itemize}

We also explain the significance of the shape operator when the geometry of the ambient space that the surface is embedded in is non-Euclidean. Under the definitions of isotropic geometry, the formula for the shape operator of a height map simply becomes the Hessian of the height function. 
From this perspective, minimizing our energy using the isotropic model can be seen as approximately minimizing the height function's interpolation error over the mesh.

\section{Related Work}

\subsection{Surface meshing}
One of the most significant results influencing the field of surface meshing is the optimality of the Delaunay triangulation.
Specifically, Rajan \cite{10.1145/109648.109688} found that the Delaunay triangulation minimizes the maximum bound on interpolation error over a domain when the vertices are fixed. 
Chen and Xu \cite{chen04} proposed an energy whose minimizer is the Optimal Delaunay Triangulation (ODT) when the vertex locations are freed. 

The Delaunay triangulation has been generalized to the anisotropic case for use in surface meshing in many ways. 
Fu \etal \cite{fu14} modified the ODT energy for use in anisotropic surface meshing and used the Hessian of the linear interpolation error to determine the anisotropy of each element. 

Another approach is to find the dual of a Voronoi diagram under a prescribed metric \cite{10.1145/777792.777822}. 
Specifically, the Centroidial Vornoi Diagram (CVT) is desirable for producing high quality meshes; a version of it (later improved by \cite{XIAO201852}) can be seen as dual to the ODT \cite{10.1145/2980179.2980245}. 
The anisotropic CVT was first adapted to surface meshing by Du and Wang \cite{doi:10.1137/S1064827503428527}. L{\'e}vy and Liu \cite{levy:inria-00600251} and Nivoliers \etal \cite{nivoliers:hal-01202738} further developed the method, using the variance of the normals to determine the anisotropy. Similar ideas were explored by Richter and Alexa \cite{RICHTER201548} and Cai \etal\cite{cai2018surface}.

An anisotropic particle-based approach was taken by Zhong \etal \cite{10.1145/2461912.2461946}. The method first optimizes a distribution of particles, then computes an anisotropic Voronoi diagram and attains the dual triangulation.

For a more complete review of remeshing techniques, we refer the reader to Khan \etal \cite{9167456}.

\subsection{Moving mesh methods}
Moving Mesh methods \cite{huang11, HUANG2001383, CAO1999221, doi:10.1137/S1064827503428527, BUDD1999756} are a set of methods typically used to generate meshes for interpolating data or solving PDEs. In contrast to other mesh generation methods, they iteratively improve an initial mesh without changing its topology. In the one-dimensional case, the method is given a prescribed density function over a domain and an existing discretization; it then attempts to adjust the distribution of points to match a desired density distribution.

This scheme can be generalized to two (and higher) dimensions by replacing the scalar-valued density function with a matrix-valued \emph{monitor function}. The monitor function is a symmetric, positive-definite tensor defined over the domain that describes the principal directions and magnitudes of stretch that the mesh should have at each point. As proposed by Dvinsky \cite{DVINSKY1991450} and later developed by others \cite{LI2001562, CHARAKHCHYAN1997385, AZARENOK2006720, AZARENOK20071102}, moving mesh methods can be generalized to use monitor functions by minimizing a Dirichlet energy measured using a metric given by the monitor function. Minimizers of this energy can be seen as discrete analogs to \emph{harmonic maps}.

Despite the long history of these techniques, there has been little research on applying moving mesh methods to work on curved surfaces. Crestel \etal\cite{CRESTEL2015148} adapted moving mesh methods to the case of parametric curved surfaces. The authors even note in the introduction that, to their knowledge, there are no published articles adapting moving meshes to surfaces. Unlike our method, their method generates isotropic meshes for use in solving PDEs. Their approach uses a scalar-valued prescribed density function, while we use a matrix-valued monitor function. Our method also doesn't require the surface to be parameterized.

\section{Background}

In this section, we give background on related previous efforts to characterize locally optimal triangle shapes and globally optimal triangulations for different approximation goals. We first describe the conditions for optimal scalar interpolation on flat domains, then detail similar results in the field of surface approximation.

\subsection{Optimal anisotropy for scalar interpolation}
\begin{figure}[tb]
    \centering
    \includegraphics[width=\linewidth]{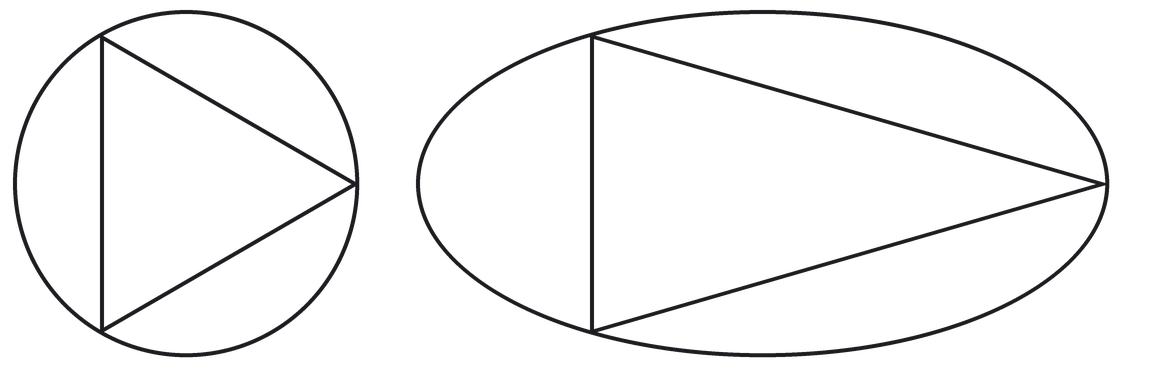}
    \includegraphics[width=\linewidth]{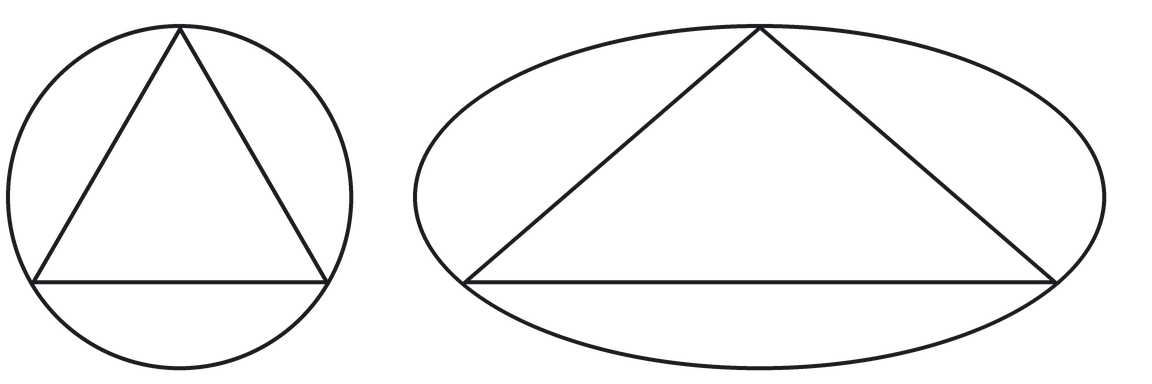}
    \caption{
    \label{fig:AnisoAB}%
        Locally optimal anisotropic triangles for scalar interpolation (right) can be generated by transforming equilateral triangles (left) using a symmetric positive definite matrix (for example, \ref{eq:OptimalLinf}). The minimal circumcircle can be stretched into the minimal circumellipse (or Steiner circumellipse) of an anisotropic triangle using the same matrix transformation. We refer to the aspect ratio of this ellipse --- which is the same as the ratios of the eigenvalues of the transformation matrix --- as the aspect ratio of a triangle. Note that while both triangles on the right may locally optimize interpolation error, the bottom anisotropic triangle might not make a good gradient approximation because it does not satisfy Shewchuck's ``no large angle'' property.
    }
\end{figure}

Some of the earliest results on anisotropic meshing focus on the local error bounds for linear interpolation on finite elements \cite{shewchuk02}. 
They specifically characterize the optimal anisotropy of an infinitesimally small triangle that locally minimizes interpolation error. In these results, the anisotropy is described by a symmetric positive-definite tensor, which is in the polar decomposition of the transformation that deforms an equilateral triangle into the anisotropic triangle.

The specific case of optimal anisotropy that minimizes \(L_\infty\) interpolation error was investigated by D'Azevedo and Simpson \cite{doi:10.1137/0912040, doi:10.1137/0910064}. They found that \(L_\infty\) optimal triangles could be calculated from the square root of the ``convexified'' Hessian of the interpolated function (\(H\)):
\begin{align}
    \label{eq:OptimalLinf}
    E^{-1} T_\Delta \textrm{, where } E = \sqrt{|H|},
\end{align}
\(T_\Delta\) is a matrix with the edges of an equilateral triangle as its column vectors, and the result is a matrix with the edges of the transformed triangle along the columns.
Here we disregard the scale of the resulting triangle for simplicity. We also define the absolute value and square root of a tensor that can be eigendecomposed into the form \(A = PDP^{-1}\) as
\begin{multline}
    \label{eq:MatAbsAndSqrt}
    |A| = P 
    \begin{bmatrix}
    |\lambda_1| & \\
     & |\lambda_2|
    \end{bmatrix}
    P^{-1}, \\
    \sqrt{A} = P 
    \begin{bmatrix}
    \sqrt{\lambda_1} & \\
     & \sqrt{\lambda_2}
    \end{bmatrix}
    P^{-1}.
\end{multline}
See Figure~\ref{fig:AnisoAB} for a few examples of anisotropic triangles.
Equivalently, one can say a triangle is optimal if it it equilateral under the metric given by \(E^2\).
In the \(L_\infty\) case this metric is simply \(|H|\).
Nadler \cite{nadler86} investigated the \(L_2\) case and Chen \etal \cite{chen07} generalized the results to \(L_p\). Under different values of \(p\), the desired triangle size relative to other triangles changes while the aspect ratio stays the same.

Typically, optimizing for linear interpolation error results in triangles with an aspect ratio of \(\sqrt{|\lambda_{max} / \lambda_{min}|}\) (where \(\lambda_{max}\) and \(\lambda_{min}\) are the maximum and minimum eigenvalues of the Hessian, respectively). However, Shewchuck \cite{shewchuk02} notes that an aspect ratio of \(|\lambda_{max} / \lambda_{min}|\) can be optimal for minimizing \(L_\infty\) \emph{gradient} error if the triangle satisfies certain requirements. We note that this aspect ratio is met if a triangle is equilateral under the metric \(|H|^2\). This can be useful as gradient error is often harder to control than interpolation error.

While these works provide a useful framework for generating locally optimal triangles, they only investigate in the context of interpolating scalar functions over a planar triangle mesh.

\subsection{Optimal anisotropy for surface approximation}

In the field of surface approximation, most anisotropic methods aim to minimize the variance in the representation's distribution of normals.
This technique was popularized by Cohen-Steiner \etal\cite{cohen-steiner04} in Variational Shape Approximation (VSA). Unlike other surface approximation methods, VSA represents a surface using a series of polygonal ``patches'' instead of a triangle mesh.

The authors argued that finding the best normal approximation can be a better goal than finding the best \(L_2\) approximation for a few reasons: first, evidence suggests our visual perception is more sensitive to changes in normals; second, (as noted by Shewchuck in the previous section) gradient error is much more difficult to control than interpolation error. VSA uses a metric similar to shape operator squared, minimizing (for a given patch \(i\) covering a region \(R_i\))
\begin{align}
    \int_{x \in R_i} \|n(x)-n_i\|^2 =  \int_{x\in R_i} \|p(x) - p_i\|_{S^2}^2.
\end{align}
This objective function is geometrically meaningful: it (heuristically) finds an even distribution of patches on the Gauss image of the surface. 

Interestingly, the aspect ratio of the elements generated using this metric mirror the results in the previous section. The normal variance metric results in elements with an asymptotic aspect ratio of $|k_{max} / k_{min}|$, while the asymptotic ratio for elements generated using the $L_2$ position variance metric is $\sqrt{|k_{max} / k_{min}|}$.

Clarkson \cite{clarkson06} proved that Hausdorff optimal surface triangulations are Delaunay triangulations of Delone (or near-uniform) sets under a metric given by the convexified shape operator. We note that this mirrors the \(L_\infty\) bound on interpolation error, which uses the convexified Hessian as a metric.

\section{Adaptive Harmonic Mappings}

In the following sections we introduce our adaptive surface meshing energies and their respective gradient flows. These energies are formulated using smooth maps; we apply these energies to the discrete case of triangle meshes in the implementation section.

We first describe a general energy over abstract manifolds, then illustrate concrete examples of this energy on one dimensional and two dimensional embedded surfaces.

\subsection{General theory}
The continuous version of the adaptive surface meshing problem can be formulated as follows. 
Let \(M\) be an \(n\)-dimensional manifold representing a \emph{target manifold}.
\(M\) can be represented by the zero level set of a scalar function defined over the space \(M\) is embedded in; in \(\RR^3\), this can be a signed distance function (SDF).
The target manifold is equipped with a metric \(\sigma\in \Gamma(T^*M\otimes_{\mathrm{sym}}T^*M)\) which controls the desired density and anisotropy for the mesh.
Now, we can describe a mesh over \(M\) as a map \(\varphi\colon\Mhat\to M\) from a \emph{material space} \(\Mhat\), which is a Riemannian manifold of the same dimension as \(M\).  Let \(g\in\Gamma(T^*\Mhat\otimes_{\mathrm{sym}}T^*\Mhat)\) denote the Riemannian metric on the material space \(\Mhat\).
If \(f\colon M\rightarrow\RR^3\) is an embedding of \(M\) in 3D, then we let \(\phi = f\circ \varphi\) denote the mesh \(\varphi\) using the 3D coordinates.

\begin{definition}
Given \((\Mhat,g)\) and \((M,\sigma)\), the \emph{Dirichlet energy} of a map \(\varphi\colon\Mhat\to M\) is defined as
\begin{align}
    \label{eq:DirichletEnergy}
    \cE(\varphi)\coloneqq {1\over 2}\int_M \sigma\left(d\varphi\mathbin{\wedge\star_g}d\varphi\right).
\end{align}
\end{definition}

Here, the integrand \(\sigma\left(d\varphi\mathbin{\wedge\star_g}d\varphi\right)\in\Omega^n(M;\RR)\) is parsed as that for any \(g\)-orthonormal basis \(e_1,\ldots, e_n\in T_p\Mhat\) at any point \(p\in\Mhat\),
\begin{align}
    \sigma\left(d\varphi\mathbin{\wedge\star_g}d\varphi\right)\big\vert_{p}(e_1,\ldots,e_n) = \sum_{i=1}^n\sigma\left(d\varphi(e_i),d\varphi(e_i)\right).
\end{align}

\begin{problem}
\label{problem1}
With \((M,\sigma)\) and \((\Mhat,g)\) given, we find a surjective map \(\varphi\colon\Mhat\to M\) that solves
\begin{align}
    \underset{\varphi\colon\Mhat\to M}{\mathrm{minimize}}\,\,\cE(\varphi).
\end{align}
Each critical point of this minimization problem is called a \emph{harmonic map}.
\end{problem}

Some of the most important results on harmonic maps, including the first variation of (\ref{eq:DirichletEnergy}), were developed by Eells and Sampson \cite{10.2307/2373037}. As a result, we know that the stationary point of this energy satisfies the following for every \(\mathring{\varphi}\):
\begin{align}
    \sigma(\mathring{\varphi},\,{}\star_g d^{\nabla_\sigma}\star_g d\varphi) = 0,
\end{align}
where \(\star_g\) denotes the Hodge star on metric \(g\) and \(d^{\nabla_\sigma}\) denotes the exterior covariant derivative using the Levi-Civita connection \(\nabla_\sigma\) on \(\varphi^*TM\), compatible with \(\sigma\). 
The functional gradient of this energy w.r.t \(\sigma\), given by \(\nabla\cE(\varphi) = {}\star_g d^{\nabla_\sigma}\star_g d\varphi\), is often referred to as the \emph{tension field}.

One can expand the exterior covariant derivative to make the connection more explicit, exposing the Laplacian plus a term involving the \emph{connection one form}.
Under any coordinate system for points on \(M\), one can expand the exterior covariant derivative \(d^{\nabla_\sigma}\) explicitly as \(d^{\nabla_\sigma}=d+\omega_{\,\sigma}\wedge{}\) involving the componentwise exterior derivative \(d\) and a matrix of one-forms \(\omega_\sigma\).
This yields the componentwise Laplacian plus a transformation of \({}\star_g d\varphi\) that accounts for how the coordinate twists and stretches along the surface:
\begin{align}
    {}\star_g d^{\nabla_\sigma}\star_g d\varphi = {}\star_g d\star_g d\varphi + \omega_{\,\sigma}\mathbin{\wedge\star_g}d\varphi.
\end{align}
To gain a better intuition of harmonic maps, one can imagine a harmonic map as a map that minimizes the distance from each point on the target manifold to the the points around it. Eells and Sampson made this intuition rigorous, showing that \(\cE\) can be approximated to the second order as:
\begin{multline}
    \label{eq:SecondOrderApprox}
    {2\epsilon^2\over n+2}+\cE\vert_p(\varphi)+O(\epsilon^3) = \\{1\over |B_\epsilon(p)|}\int_{B_\epsilon(p)}d_\sigma^2\bigl(\varphi(p), \varphi(q)\bigl)\ \mu_g(q),
\end{multline}
where \(B_\epsilon(p)\) is a ball in \(\Mhat\) with radius \(\epsilon\) centered at \(p\) with volume given by \(|B_\epsilon(p)|\), \(d_\sigma^2(\cdot, \cdot)\) is the distance squared between two points measured using \(\sigma\), and \(\mu_g(q)\) is a volume form on \(\Mhat\) using metric \(g\) at point \(q\). Interestingly, this is a surprisingly powerful property: Jost \cite{Jost1994EquilibriumMB} used this approximation to generalize harmonic maps to functions between two metric spaces. We show that in the discrete case of triangle meshes, our method minimizes a similar energy.

\subsection{Mesh adaptation in one dimension}
\label{sec:MeshAdaptationIn1D}
The general energy introduced above can be better physically understood when applied to a one-dimensional mapping. Imagine the surface we would like to represent is a unit length parameterized curve given by \(f\colon I\to\RR^3\) defined over an interval \(I\subset\RR\).

Let the map \(\varphi\colon \widehat{I}\to I\) deform a rod bent into the shape of the curve by moving the material at location \(x\in \widehat{I}\) along the curve to a new point at location \(\varphi(x)\in I\) along the curve.
If \(\varphi\) is a minimizer of \(\cE(\varphi)\), then it has stretched and compressed the rod in such a way that it has uniform density under \(\sigma\), as we show below.

Let the curvature of a point at length \(t\) along the curve be given by \(\kappa(t) > 0\). If we set \(\sigma\vert_t(\cdot, \cdot) = \kappa^2(t)\bigl((\cdot)(\cdot)\bigl)\), then the energy and its variation can be explicitly written as:
\begin{subequations}
\begin{align}
    \cE_{1D}(\varphi) &= \int_{\widehat{I}}(\kappa\,\varphi')^2, \\
    \mathring\cE_{1D}(\varphi) &= \int_{\widehat{I}}\kappa\,\mathring\varphi\,(\kappa\,\varphi')'.
\end{align}
\end{subequations}
Alternatively, the variation can be written as \(\kappa^2\,\mathring\varphi\,(\varphi'' + \varphi'{\kappa'/\kappa})\) to expressly reveal the covariant derivative term seen in the variation of the general energy.

We now show that the rod's density \(\rho = 1/\varphi'\propto\kappa\) when \(\cE_{1D}\) is minimized. The Euler-Lagrange equation is \((\kappa\,\varphi')'=0\), which tells us \(\kappa\,\varphi'=c\) for some \(c\). Re-arranging the equation, we find that \(\rho\propto\kappa\).

Currently, this method of manipulating \(\varphi\colon\widehat{I}\to{I}\) requires an intrinsic coordinate for \(I\).
In practice, we would like to work with the 3D coordinates \(\phi = f\circ\varphi\colon\widehat{I}\to\RR^3\) directly.
After a change of variable, we write
\(\cE(\phi)\) and its variation as:
\begin{subequations}
\label{eq:Energy1DExtrinsic}
\begin{align}
    \cE_{1D}(\phi) &= \int_{\widehat{I}}\|\kappa\,\phi'\|^2, \\
    \mathring\cE_{1D}(\phi) &= \int_{\widehat{I}}\left\langle\kappa\,\mathring\phi, \cP_{T}\bigl((\kappa\,\phi')'\bigl)\right\rangle_{\RR^3},
\end{align}
\end{subequations}
where \(\cP_{T}\) projects vectors in \(\RR^3\) to the tangent of the curve. 
Note that the intrinsic covariant derivative, which will be more complicated in 2D, has a simpler form in the extrinsic coordinate with the aid of a projection.  We will also utilize the extrinsic coordinate in Section~\ref{sec:MeshAdaptationIn2D}.

\subsection{Mesh adaptation in two dimensions}
\label{sec:MeshAdaptationIn2D}
We now apply the general energy (\ref{eq:DirichletEnergy}) to a two-dimensional mapping. Let \(\varphi\) now be a map from a two dimensional space \(\Mhat\) to \(M\) that deforms a material of uniform density in \(\Mhat\).
Similarly to the treatment in \eqref{eq:Energy1DExtrinsic}, we represent the map \(\varphi\) using the 3D coordinates \(\phi = f\circ\varphi\) where \(f\colon M\to\RR^3\) is \(M\)'s embedding into 3D.
The embedding \(f\) induces a metric \(h\)
and a shape operator \(S\).

Let \(|S|\) be the symmetric positive definite ``convexified'' shape operator, using the absolute value of a matrix defined in (\ref{eq:MatAbsAndSqrt}) and with an addition of a small amount of the identity if necessary.
Building on the example of Section~\ref{sec:MeshAdaptationIn1D}, it is natural to consider \(\sigma(\cdot, \cdot) = h(|S|(\cdot), |S|(\cdot))\). In this case, the energy and its variation can be written as:
\begin{align}
    \cE_{2D}(\phi) &= \int_{\Mhat}h(|S|\,d\phi\mathbin{\wedge\star}|S|\,d\phi), \\
    \mathring\cE_{2D}(\phi) &= \int_{\Mhat}h(|S|\,\mathring\phi,\ |S|\star d^{\nabla_\sigma}\star d\phi).
\end{align}
This energy shares many similarities with the well-studied Dirichlet energy, which is typically used in applications like heat diffusion and mean curvature flow. On regions where the shape operator \(S\) is positive definite, the energy can be interpreted as the Dirichlet energy of the induced Gauss map \(n\circ \varphi\colon \Mhat\to\SS^2\) where \(n\colon M\to\SS^2\) is the surface normal.

In practice, we rewrite the gradient \(\nabla\cE_{2D}(\phi) = {}\star d^{\nabla_\sigma}\star d\phi\) of this energy to a simpler form that uses the Levi-Civita connection \(\nabla_h\) compatible with the metric \(h\). Using the property \(|S|\,d^{\nabla_\sigma}\eta = d^{\nabla_h}(|S|\,\eta)\) for an arbitrary tangent-vector-valued \(k\)-form \(\eta\),
\begin{align}
    \nabla\cE_{2D}(\phi) = {}\star d^{\nabla_h}\star |S|\,d\phi.
\end{align}
Conveniently, in \(\RR^3\) the covariant derivative is the tangent space component of the gradient of a vector field. Therefore, \(\nabla\cE_{2D}\) can be further re-written as:
\begin{align}
    \nabla\cE_{2D}(\phi) = \cP_{TM}({}\star d\star |S|\,d\phi).
\end{align}
Our 1D formula \eqref{eq:Energy1DExtrinsic} for the variation can be derived though similar reasoning.

Eells and Sampson showed that if \(\phi\) is a harmonic function, \(d\phi\) is a harmonic tangent-vector-valued one form satisfying \(\delta^{\nabla_\sigma}\,d\phi = 0\) and \(d^{\nabla_\sigma}d\phi = 0\) (or equivalently, \(\delta^{\nabla_h}\,|S|\,d\phi = 0\) and \(d^{\nabla_h}|S|\,d\phi = 0\)). Let \(\gamma\) be a harmonic tangent-vector-valued one form under the connection \(\nabla_h\). Then --- similarly to how \(\kappa\,\varphi' = c\) in the 1D example --- we have \(|S|\,d\phi = \gamma\) in 2D. 

Interestingly, \(\gamma\) may have some anisotropy that accounts for overall ``stretch'' in the domain. For example, imagine a harmonic map from a square domain to a rectangle using their respective standard Euclidean metrics. Or, let the target metric be non-standard --- imagine the target manifold is now a square domain with a metric induced from a map stretching the square into a rectangle. This hints at a limitation in our method and moving mesh methods in general: depending on \(\sigma\) and the initial distribution of mass in the material space, solutions to \probref{problem1} may have some leftover anisotropy in the density distribution under \(\sigma\).

\subsection{Mesh adaptation in \(\II^3\)}

Producing meshes for interpolating height fields (or more generally, scalar functions) over flat domains has been studied extensively by moving mesh literature.
In our adaptive surface meshing method, by switching the geometry of the ambient space from Euclidean to \emph{isotropic}, we obtain an adaptive surface meshing algorithm for height fields.
In other words, our work can be seen as an alternative generalization of previous works. We note that the word ``isotropic'' in this case is not to be confused with the notion of isotropy vs. anisotropy in meshing.

Isotropic geometry, developed by Austrian geometer Karl Strubecker in the 1930s and 40s, is a non-Euclidean geometry which has shown useful for describing the geometry of function graphs analogous to the differential geometry of surfaces embedded in a Euclidean space. The results we need are covered by Pottmann and Liu in \cite{10.1007/978-3-540-73843-5_21}.

The isotropic metric is given solely using the \(x\) and \(y\) coordinates, as if measuring from a top down view. In the isotropic space, the ``unit sphere'' in Euclidean geometry is replaced by the paraboloid \(z = {1\over 2}(x^2 + y^2)\).  In particular, the Gauss map of a surface \(\{(x,y,f(x,y))\}\) in isotropic geometry maps each point of the surface to a point on the paraboloid whose tangent plane is parallel to the tangent plane at the original surface point.
The formula for the Gauss map can be given by:
\begin{align}
n: (x, y, f(x, y)) \mapsto (f_x, f_y, \frac{1}{2}(f_x^2 + f_y^2)).
\end{align}
Taking the gradient of this map, we obtain the isotropic shape operator; taking the determinant of the shape operator, we find the isotropic Gauss curvature:
\begin{align}
    \nabla n = \nabla^2 f, ~~~~
    K = \textrm{det}(\nabla^2 f).
\end{align}
Therefore, a minimizer of \(\cE_{2D}\) in isotropic space would have a uniform density distribution under the metric \(H^2\) given by the Hessian \(H\) of \(f\). We cover the relationship between the Hessian and the shape operator further in the next section.

\subsection{Controlling anisotropy}

In many cases it is useful to provide a way for the user to control the anisotropy of the resulting triangles. We slightly modify \(\cE_{2D}\) by introducing an anisotropy parameter \{\(\alpha\in\RR\ |\ 0\leq\alpha\leq 1\)\} to \(\sigma\) in \(\cE_{2D}\):
\begin{align}
    \cE_{2D,\,\alpha}(\phi) = \int_{\Mhat} h(|S|^\alpha d\phi\mathbin{\wedge\star}|S|^\alpha d\phi),
\end{align}
where \(A^\alpha\) denotes the matrix power. We now analyze the properties of meshes consisting of near-equilateral triangles of equal area under \(\sigma\) with different \(\alpha\). 

\medskip
\noindent
\textbf{Effect of \(\alpha\).}\,\,
If \(\alpha=1\), then triangles are near-equilateral on the Gauss image. We also note that the distribution of vertex normals is approximately uniform, so minimizing this energy heuristically approaches a triangulation where the normals do not vary much over each face. In the isotropic case, this intuition has been made rigorous by Shewchuck \cite{shewchuk02}: equilateral triangles under this metric meet the aspect ratio criteria required to become ``super-accurate'' gradient approximations. However, we are careful to note that they are not guaranteed to meet the other criteria listed.

If \(\alpha=0.5\), triangles become roughly equilateral under the metric \(|S|\). In Euclidean space, the mesh becomes a near-optimal Hausdorff approximation of the surface. As Clarkson \cite{clarkson06} showed, the Delaunay triangulation of a "uniform" set of points (or more specifically, a Delone set) under the metric \(|S|\) is Hausdorff optimal. This is mirrored in the isotropic case, where equilateral triangles under \(|H|\) minimize the local \(L_\infty\) interpolation error of \(f\).

If \(\alpha=0\), then the matrix power degenerates to \(I\) and minimizing \(\cE_{2D}\) produces near-equilateral triangles.

This modification can be adapted to the one dimensional energy in the following way:
\begin{align}
    \cE_{1D,\alpha}(\phi) = \int_{\widehat{I}}(\kappa^\alpha~\phi')^2.
\end{align}

\section{Implementation}

In this section, we describe how the smooth energies of the previous section are adapted to the discrete case of triangle meshes.
A Houdini implementation of the algorithm can be found here: \url{https://nickwn.github.io/asmhm_supp.zip}. 

\subsection{Discrete 2D energy approximation}

In the discrete formulation, we would like to minimize \(\cE_{2D}\) over the map \(\phi\) from a 
triangle mesh \(\Mhat\) to a given surface \(M\).
The metric \(g\) on \(\Mhat\) is chosen so that every triangle is equilateral and equal-area.
The deformation of a triangle on \(M\) is measured using the metric \(\langle|S|(\cdot), |S|(\cdot)\rangle_{\RR^3}\) which operates on vectors in \(\RR^3\) and is piecewise constant over each triangle. 
\(|S|\) can be extended to an operation on vectors in \(\RR^3\) by pre-projecting vectors to \(TM\).

Because the triangles on \(\Mhat\) are all equal-area and equilateral, the Hodge star \(\star_g\) on \(\Mhat\) for one forms and two forms can be factored out as a constant term and ignored. In the case of one forms, the discrete Hodge star calculated using the cotangent formula \cite{em/1062620735} is the same on all edges of the mesh because all the angles are the same. In the case of two forms, the discrete Hodge star is constant because the area of the dual cell for each point is also constant.

Let \(\cE_{2D}\vert_i\) denote the discrete energy at a vertex \(i\). Then, using the above insights, we can approximate the discrete energy at vertex \(i\) as:
\begin{align}
    \label{eq:DiscreteEnergy}
    \cE_{2D}\vert_i(\phi) = \sum_{j\in {\rm Star}(i)}\||S|\,d\phi_{ij}\|^2,
\end{align}
where \(d\phi_{ij}\) denotes the discrete differential of \(\phi\) along edge \(ij\). We are intentionally vague describing which face \(|S|\) is defined on because \(\||S|_{ijk}d\phi_{ij}\| = \||S|_{jkl}d\phi_{ij}\|\). This mirrors the smooth ``averaging'' approximation (\ref{eq:SecondOrderApprox}); in Figure~\ref{fig:EnergyGraph}, we show that our gradient descent approximation heuristically minimizes this discrete energy.

\subsection{2D projected backward Euler solver}
\label{sec:2DFlow}
\definecolor{offblack}{rgb}{0.45, 0.4, 0.5}
\begin{figure}[tb]
\begin{tikzpicture}
\begin{axis}[
    xlabel={Iteration},
    ylabel={Discrete \(\cE_{2D}\)},
    ymajorgrids=true,
    grid style=dashed,
    width=\linewidth,
    height=5cm
]
\addplot[line width=0.5mm, color=offblack] table {energyplot.dat};
\end{axis}
\end{tikzpicture}
\includegraphics[width=\linewidth]{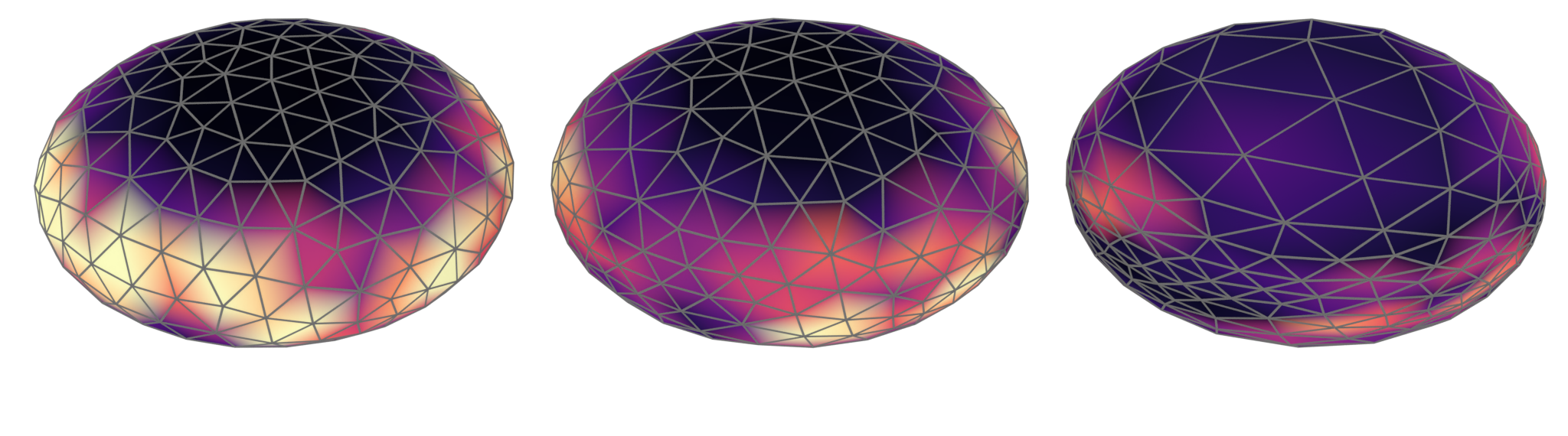}
\includegraphics[width=0.5\linewidth]{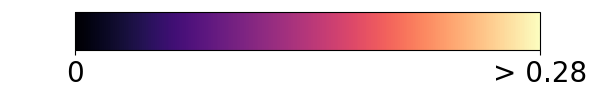}
\caption{
        \label{fig:EnergyGraph}
        Discrete \(\cE_{2D}\) energy (\ref{eq:DiscreteEnergy}) (top) measured on an ellipsoid over 50 iterations of (\ref{eq:2DGradientDescent}), with \(\alpha = 1\) and \(\epsilon = 0.03\). Ellipsoids shown (bottom) are at 0 (left), 5 (middle), and 50 (right) iterations, where the color visualizes \(\det(|S|d\phi)\). Note that \(\det(|S|d\phi)\) becomes uniform as the iteration progresses.
    }
\end{figure}
\begin{figure}[tb]
    \centering
    \makebox[\linewidth][c]{\includegraphics[width=\linewidth]{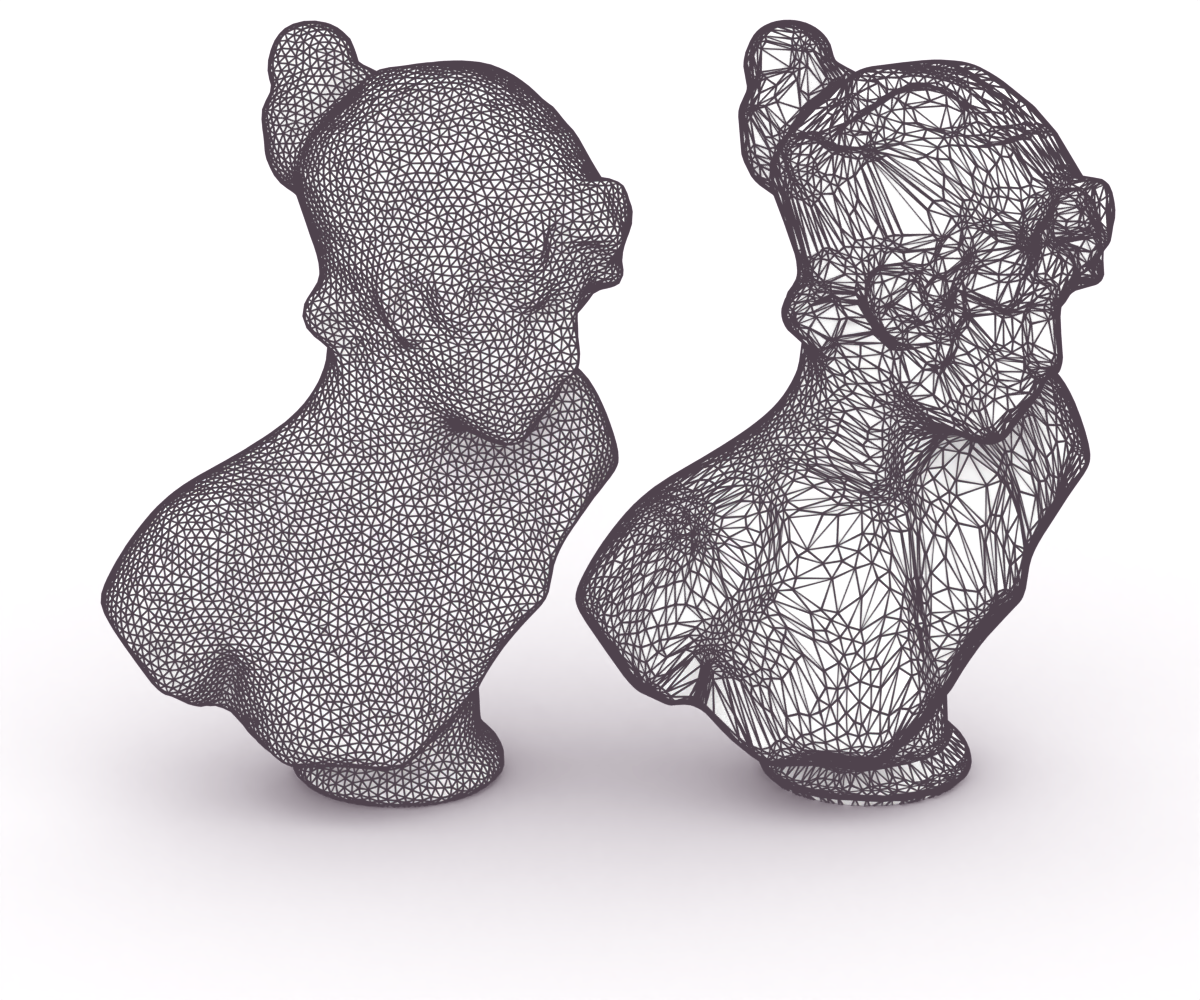}}
    {\includegraphics[width=\linewidth]{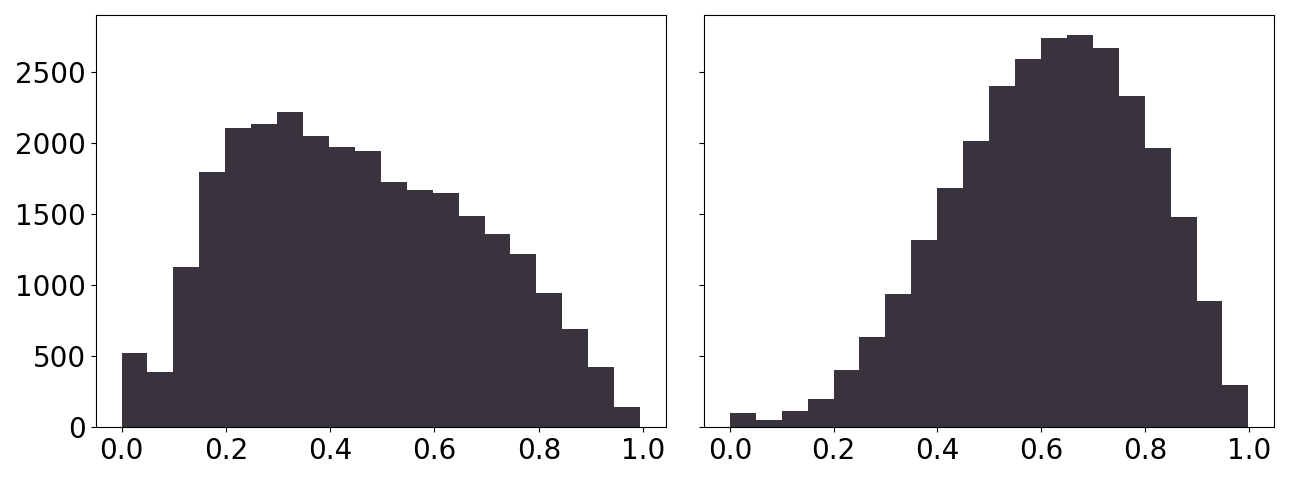}}
    \caption{
    \label{fig:Bust}%
        2D approximation (\ref{eq:2DGradientDescent}), with \(\alpha=0.75, \epsilon = 0.01\) at 0 (left) and 500 (right) iterations. 
        Histograms of the triangle aspect ratios (measured under \(\sigma\)) for each mesh are also shown below. Note that our method doesn't converge to a distribution of aspect ratios perfectly centered around 1; instead, the center is slightly less because of the harmonic ``leftover'' anisotropy described in Section~\ref{sec:MeshAdaptationIn2D}.
    }
\end{figure}
\begin{figure}[tb]
    \centering
    \makebox[\linewidth][c]{\includegraphics[width=1.1\linewidth]{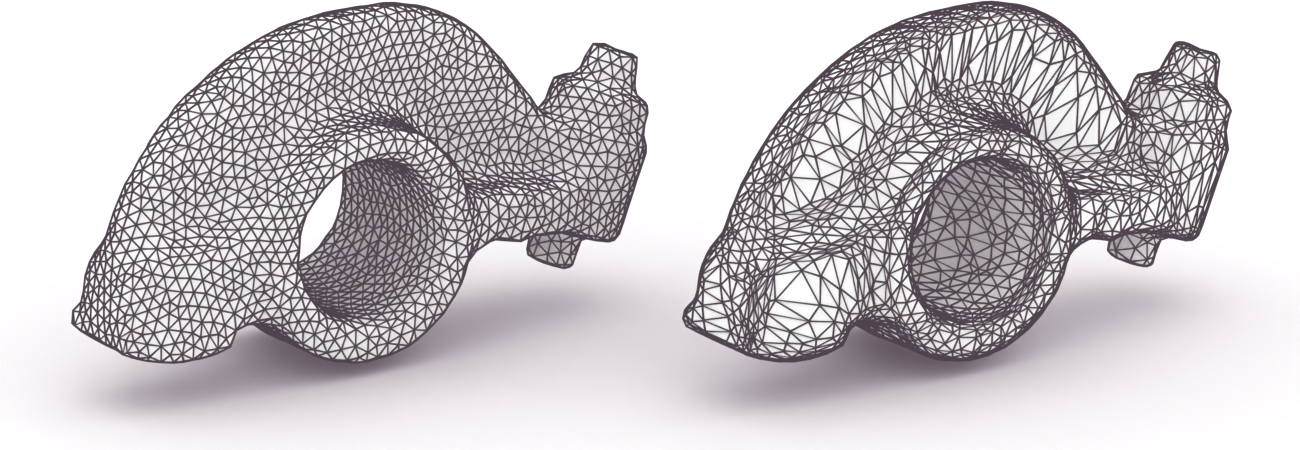}}
    {\includegraphics[width=\linewidth]{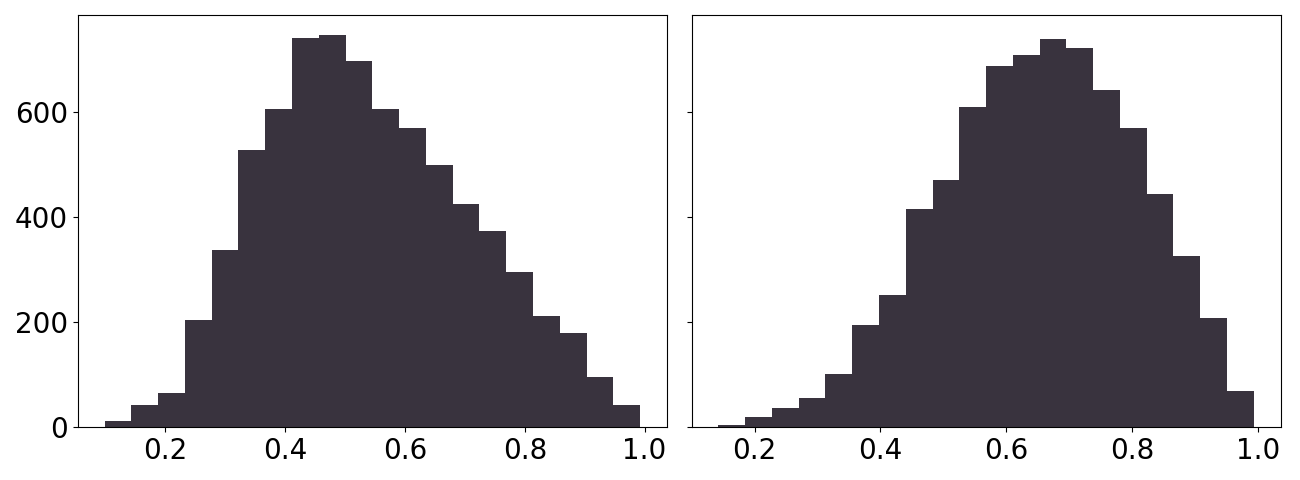}}
    \caption{
    \label{fig:Rocker}%
        2D approximation (\ref{eq:2DGradientDescent}), with \(\alpha=0.5, \epsilon = 0.01\) at 0 (left) and 500 (right) iterations. Histograms of the triangle aspect ratios (measured under \(\sigma\)) for each mesh are also shown below.
    }
\end{figure}
We now describe our method to approximate gradient descent on \(\cE_{2D}\). Building on the spatial discretization described in the previous section, we discretize the diffusion process \(\dot\phi = \cP_{TM}({}\star d\star |S|d\phi)\) temporally by time-splitting a backward Euler step and a projection step.
\begin{subequations}
\label{eq:2DGradientDescent}
\begin{align}
    \phi^{t+0.5} &= \phi^t+\epsilon\ d^\intercal\,|\bS|\,d\,\phi^{t+0.5}  \\
    \phi^{t+1} &= \textbf{P}_Z(\phi^{t+0.5})
\end{align}
\end{subequations}
In the above, \(\phi\) is a \(3V \times 1\) column vector with the \(xyz\) components flattened, \(d\) is a \(3V \times 3E\) matrix calculating the discrete difference of each \(xyz\) component along each edge in the mesh, and \(|\textbf{S}|\) is a \(3E \times 3E\) matrix that transforms each \(d\phi^{t+0.5}_{ij}\) by the sum of the shape operators on each adjacent face --- that is, it has \(|S|^t_{ijk} + |S|^t_{jkl}\) along the diagonal for each edge. Similarly to (\ref{eq:DiscreteEnergy}) in the previous section, the discrete Hodge stars have been factored out.

 \(\textbf{P}_Z\) is a function that projects points in \(\RR^3\) to the zero level set of the signed distance function \(SDF\). \(\textbf{P}_Z\) can be given by:
\begin{align}
    \textbf{P}_Z(\phi_i) := \phi_i - \nabla\bigl(SDF^2(\phi_i)\bigl).
\end{align}
The term \(d^\intercal\,|\bS|\,d\) on a triangle mesh at a vertex \(i\) can be more clearly written as the sum of the tensions from both triangles adjacent to each edge.
\begin{subequations}
\begin{multline}
    d^\intercal\,|\bS|\,d\phi^{t+0.5}\ \vert_i = \sum_{j \in{\rm Star}(i)} (|S|_{ijk}^t + |S|_{jkl}^t)d\phi_{ij}^{t+0.5} \\
    = \sum_{j \in{\rm Star}(i)} |S|_{ijk}^td\phi_{ij}^{t+0.5} + |S|_{jkl}^td\phi_{ij}^{t+0.5}
\end{multline}
\end{subequations}
In practice, the backward Euler step can be written as a linear equation solve:
\begin{align}
    (I - \epsilon~d^\intercal\,|\bS|\,d)\phi^{t+0.5} = \phi^t.
\end{align}
We show how the discrete energy (\ref{eq:DiscreteEnergy}) in the previous section converges after multiple iterations of (\ref{eq:2DGradientDescent}) in Figure~\ref{fig:EnergyGraph}.
See Figures \ref{fig:Bust} and \ref{fig:Rocker} for examples showing how the flow performs on more complex surfaces.

\subsection{1D Spring system approximation}
\label{sec:1DFlow}
\begin{figure*}[tb]
    \centering
    \mbox{} \hfill
    \includegraphics[width=\linewidth]{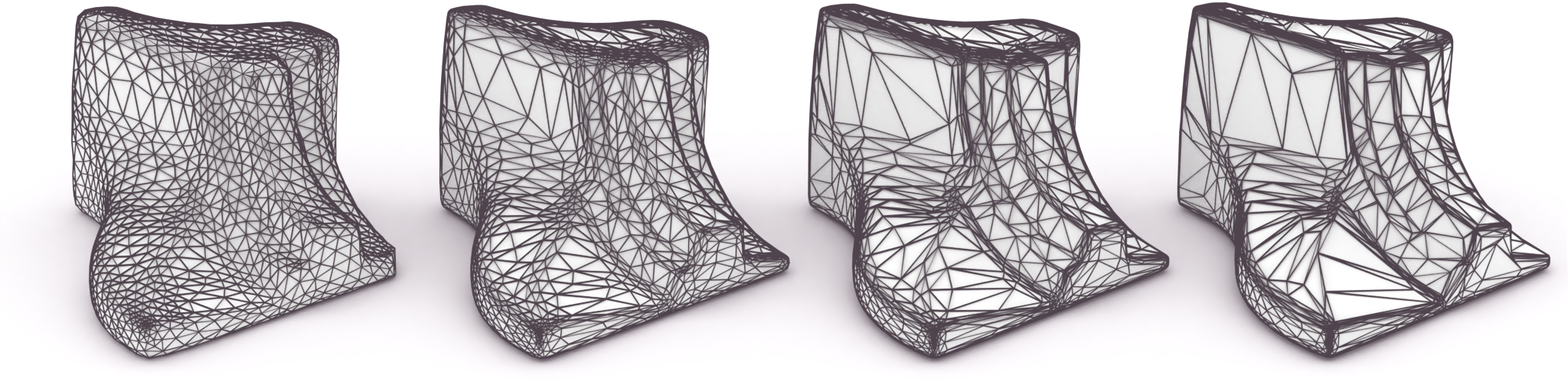}
     \hfill \mbox{}
    \caption{
    \label{fig:Aniso2D}%
        2D approximation (\ref{eq:2DGradientDescent}), with \(\alpha=0.25, 0.5, 0.75,\ \textrm{and}\ 1.0\) (left to right).
    }
\end{figure*}
\begin{figure*}[tb]
    \centering
    \mbox{} \hfill
    \includegraphics[width=\linewidth]{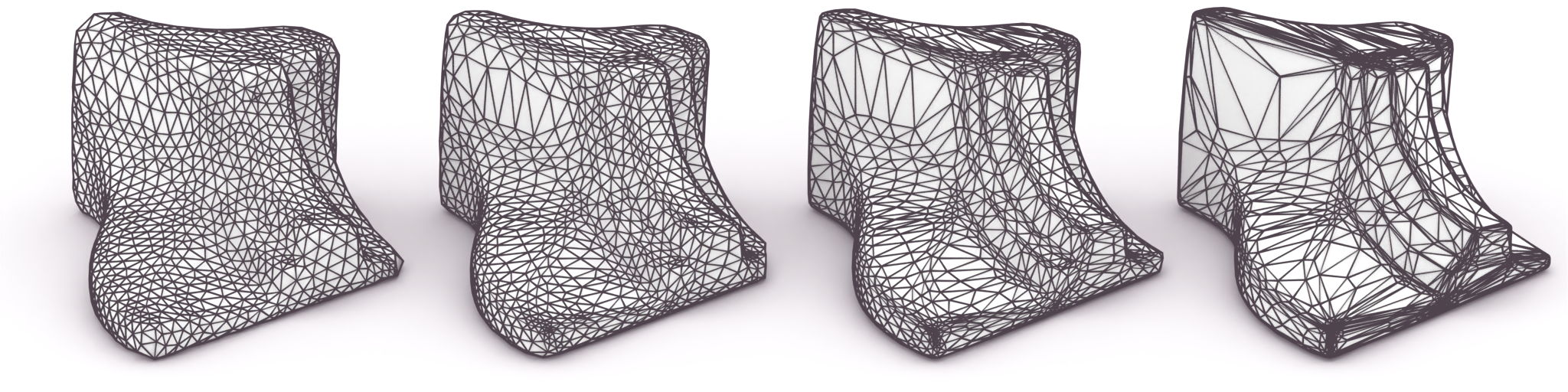}
     \hfill \mbox{}
    \caption{
    \label{fig:Aniso1D}%
        1D spring graph approximation (\ref{eq:1DGradientDescent}), with \(\alpha=0.25, 0.5, 0.75,\ \textrm{and}\ 1.0\) (left to right).
    }
\end{figure*}
\begin{figure*}[tb]
    \centering
    \mbox{} \hfill
    \includegraphics[width=\linewidth]{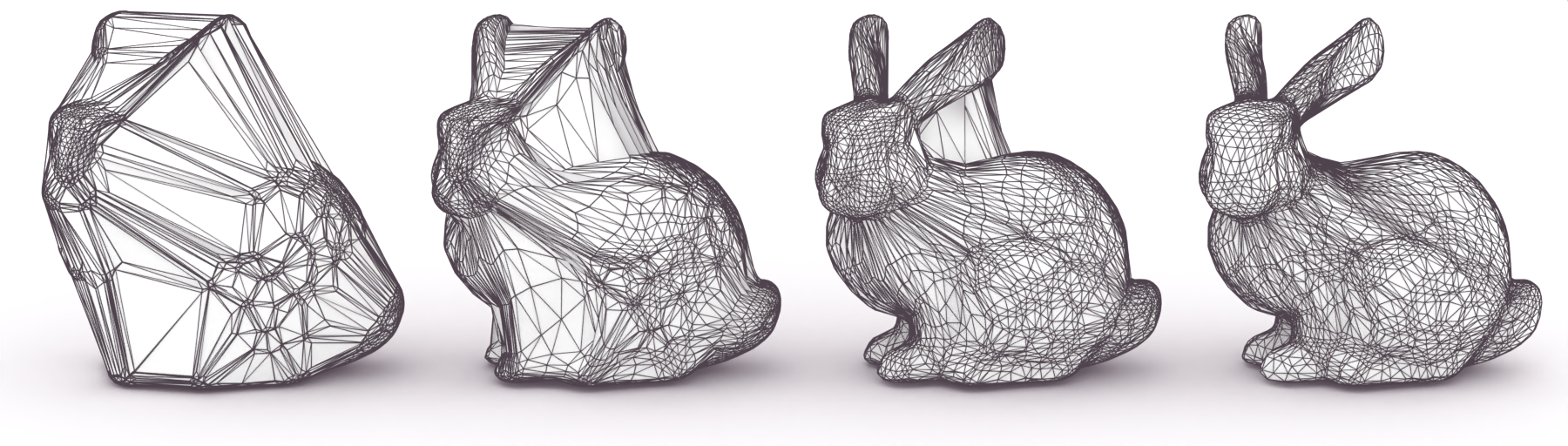}
     \hfill \mbox{}
    \caption{
    \label{fig:BadBunny}%
        We improve a sphere mesh projected to a bunny using our spring graph flow (\ref{eq:1DGradientDescent}), with \(\alpha=1.0\). Bunny is shown after iterations 0, 10, 100, and 350 (left to right).
    }
\end{figure*}
While the approach in the previous section works well, it can become unstable when triangles grow too thin, causing elements to flatten completely or flip. In this section, we introduce a less formally motivated approach that is much more robust.

The previous discrete flow (\ref{eq:2DGradientDescent}) can be cheaply approximated by minimizing the one dimensional energy on the edges of a mesh. Let \(\Mhat\) and \(M\) be one dimensional manifolds topologically equivalent to the graph of the mesh. Let \(\kappa_{ij}\) be a piece wise constant scalar curvature for an edge \(ij\). Then the discrete approximation of \({}\star d\star\kappa\,d\phi\) at a vertex \(i\) can be written as:
\begin{align}
    {}\star d\star\kappa\,d\phi\ \vert_i = \sum_{j \in{\rm Star}(i)}\kappa_{ij}\,d\phi_{ij}.
\end{align}
As in the previous section, we can assume that the edges of the mesh in \(\Mhat\) measured with \(g\) are all the same length, so the discrete scalar Hodge star is constant across the mesh. In practice, however, we found that it can be useful to manipulate the discrete Hodge star on zero forms to ensure it normalizes the edge weights so the flow converges faster. This is similar to the approach used in Desbrun \etal \cite{10.1145/311535.311576} to speed up curvature flows.

The minimizer of this energy on a mesh can be physically understood as the steady state of a graph of springs, where the spring constant of each edge \(ij\) is given by \(\kappa_{ij}\) and the point mass \(m_i\) of each vertex \(i\) is given by \(\sum_{j \in{\rm Star}(i)}\kappa_{ij}\). Let \(\Tilde{\star}\) denote this modified Hodge star. Then \(\Tilde{\star}\,d\star\kappa\,d\phi\) for a vertex \(i\) can be written as
\begin{align}
    \Tilde{\star}\,d\star\kappa\,d\phi\ \vert_i = \frac{1}{\sum_{j\in {\rm Star}(i)}\kappa_{ij}}\sum_{j\in {\rm Star}(i)}\kappa_{ij}\,d\phi_{ij}.
\end{align}
Similarly to the last section, we iteratively solve for the next state by doing an implicit Euler step plus a projection step, with \(\epsilon = 1\).
\begin{align}
    \label{eq:1DGradientDescent}
    (\textbf{M} - \epsilon~d^\intercal\textbf{K}\ d)\phi^{t+0.5} = \textbf{M}\phi^t \\
    \phi^{t+1} =  \textbf{P}_Z(\phi^{t+0.5})
\end{align}
The mass matrix \(\textbf{M}\) contains \(m_i\) along the diagonal for each vertex \(i\) and \(\textbf{K}\) contains \(\kappa_{ij}\) along the diagonal for each edge \(ij\). We define \(\kappa_{ij}\) in the following section for surfaces in Euclidean and isotropic space.

This is similar to a moving mesh energy proposed by Habashi \etal \cite{Habashi00}; however, we use a shape operator-based metric and add a projection term.

We show how this 1D approximation compares to the previous 2D approximation for different values of \(\alpha\) in Figures \ref{fig:Aniso2D} and \ref{fig:Aniso1D}. As can be seen, it exhibits less anisotropy than its more accurate counterpart.
Unlike the 2D approximation, the spring graph approximation is much more robust. It performs incredibly well on initial meshes of poor quality. In Figure~\ref{fig:BadBunny}, we show how our flow behaves on a sphere mesh projected to the surface of a bunny.

\subsection{Shape operator and curvature calculation}

The 2D diffusion and spring system approximations in the previous sections only require a shape operator per triangle and scalar curvature per edge respectively. We show how to calculate these using a general isotropic or Euclidean normal known per vertex. 

 Let \(u_i, u_j\), and \(u_k\) be 2D coordinates for the vertices of a triangle \(ijk\) under a local coordinate system with basis vectors (\(x_1, x_2\)) on the tangent space of the triangle. In the isotropic case, the tangent space can be thought of as parallel the \(xy\) plane. The shape operator for a triangle \(ijk\) with vertex normals \(n_i, n_j\), and  \(n_k\) is calculated as follows. We first calculate the deformation gradient of the map from \(u\) to the Gauss image \(n\).
 \begin{align}
    U &= 
    \begin{bmatrix}
    du_{ij} & du_{ik} \\
    \vertbar & \vertbar
    \end{bmatrix}
    \in\RR^{2\times 2} \\
    N &= 
    \begin{bmatrix}
    \vertbar & \vertbar \\
    dn_{ij} & dn_{ik} \\
    \vertbar & \vertbar
    \end{bmatrix}
    \in\RR^{3\times 2}\\
    S_{3\times 2} &= N\,U^{-1}
\end{align}
Next, we extract the \(2\times 2\) symmetric part of the polar decomposition of \(S_{3\times 2}\).
\begin{align}
    S_{2\times 2} = \sqrt{S_{3\times 2}^\intercal S_{3\times 2}}
\end{align}
Let \(T\) be a matrix that projects vectors in \(\RR^3\) to the tangent basis (\(x_1, x_2\)). Then finally, we calculate \(S\) by sandwiching \(S_{2\times 2}\) with a projection to 2D (\(T\)) and a rotation back into 3D (\(T^\intercal\)).
\begin{align}
    T& = 
    \begin{bmatrix}
    \horzbar & x_1 & \horzbar \\
    \horzbar & x_2 & \horzbar \\
    \end{bmatrix} 
    \in\RR^{2\times 3} \\
    S& = T^\intercal S_{2\times 2}T
\end{align}
In practice, we also add a small amount of the identity matrix to \(|S|\) to ensure it is positive-definite.

For the spring system, the discrete curvature \(\kappa_{ij}\) along an edge connecting vertices \(i\) and \(j\) can simply be approximated by the average rate of change of the normal along the edge.
\begin{align}
    \kappa_{ij} = \frac{\|dn_{ij}\|}{\|d\phi_{ij}\|}
\end{align}
In the isotropic case, \(\|d\phi_{ij}\|\) and \(\|dn_{ij}\|\) are measured from a ``top down view'' using the ``i-distance''.

\subsection{Boundary treatment}
\begin{figure*}[tb]
    \centering
    \mbox{} \hfill
    \includegraphics[width=\linewidth]{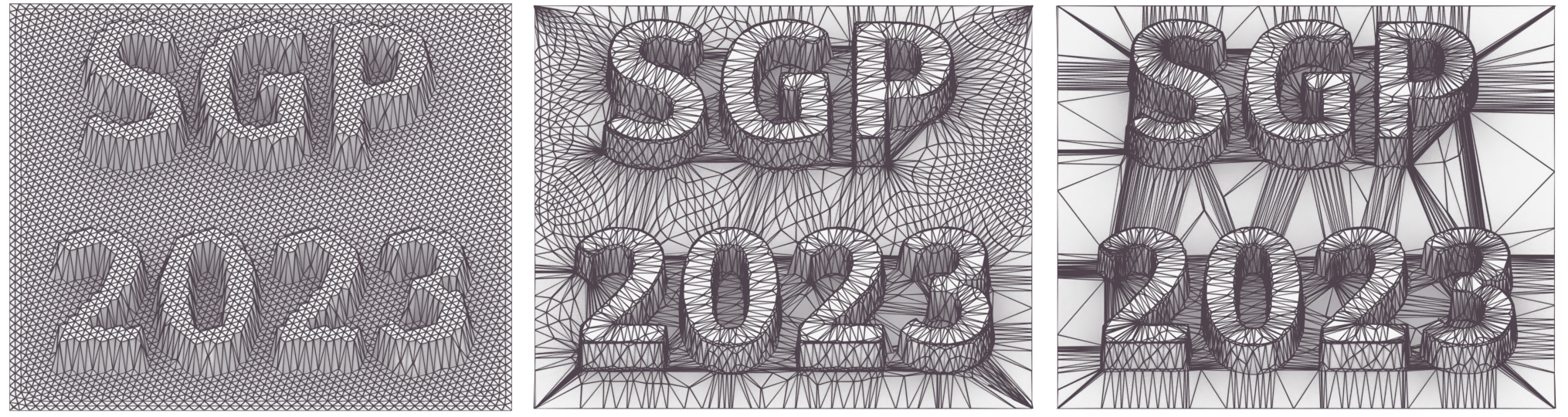}
     \hfill \mbox{}
    \caption{
    \label{fig:SGP2023}%
        Mesh of a heightmap with text after iterations 0, 25, and 125 (left to right). Uses spring graph approximation (\ref{eq:1DGradientDescent}) with \(\alpha=0.5\).
    }
\end{figure*}
\begin{figure}[tb]
    \centering
    \mbox{} \hfill
    \includegraphics[width=\linewidth]{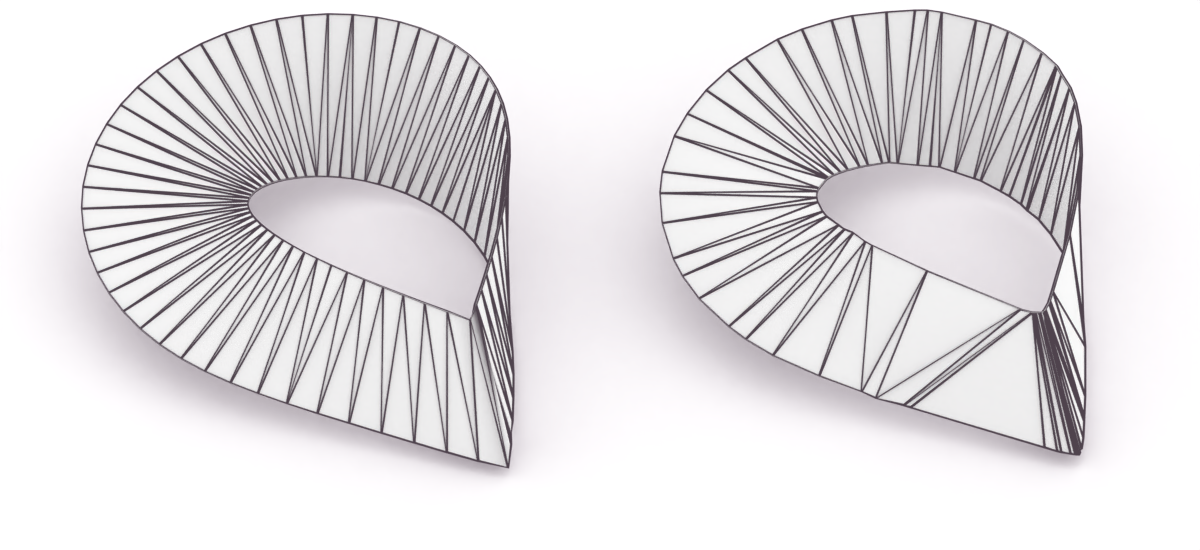}
     \hfill \mbox{}
    \caption{
    \label{fig:Mobius}%
        Mesh of a pinched Mobius strip with spring graph approximation (\ref{eq:1DGradientDescent}) with \(\alpha=1\) after 0 (left) and 200 (right) iterations. To calculate \(\kappa_{ij}\) and \((\kappa_b)_{ij}\), we chose the orientations of \(n_i\) and \(n_j\) so that they are locally consistent.
    }
\end{figure}

We treat the boundary of a mesh as a (partially) separate approximation of an underlying boundary curve, minimizing a one dimensional energy over the boundary edges of the mesh in a way that is similar to the spring system, with one exception: we set the discrete curvature of a boundary edge \(ij\) (denoted by \((\kappa_b)_{ij}\)), to the curvature of the Frenet frame along the edge. This accounts for \emph{geodesic curvature} along the boundary curve, allowing the boundary vertices to adapt to features like corners in a height map.

To calculate \((\kappa_b)_{ij}\), we construct a Darboux frame for each boundary vertex. Let \(n_i\) be the \emph{unit normal} of the Darboux frame for a vertex \(i\). In the isotropic case where the surface is given by a parametric function \(f(x, y)\), we use an alternate definition of the normal.
\begin{align}
    n_i = \frac{(f_x, f_y, 1.0)}{\sqrt{f_x^2+f_y^2+1.0}}
\end{align}
Let the \emph{tangent normal} \(t_i\) be the vector tangent to surface and orthogonal to the boundary. Then the direction of the Darboux --- and Frenet --- \emph{unit tangent} \(T_i\) is given by \(n_i\times t_i\). The curvature along an edge can then be given by the rate of change of the angle between the two unit tangents.
\begin{align}
    (\kappa_b)_{ij} = \frac{\cos^{-1}(T_i\cdot T_j)}{\|d\phi_{ij}\|}
\end{align}
Similarly to the previous section, in the isotropic case \(\|d\phi_{ij}\|\) is measured using the i-distance. 

This can be incorporated into the discrete 2D diffusion process in Section~\ref{sec:2DFlow} by setting the \(3\times 3\) entry for a boundary edge \(ij\) along the diagonal in \(|\bS|\) to \((\kappa_b)_{ij}~I_{3\times 3}\). This can be integrated into the 1D spring system formulation in Section~\ref{sec:1DFlow} by setting the entry for edge \(ij\) in \(\bK\) to \((\kappa_b)_{ij}\) and adding \((\kappa_b)_{ij}\) to the point masses \(m_i\) and \(m_j\).

In practice, we also provide a boundary weight scale factor to give the user more control over how much interior points affect the boundary. We also note that boundary points are projected back to the boundary along with interior points during the projection step. Edges between boundary points and interior points use the usual definition for \(\kappa\). A result can be seen in Figure~\ref{fig:Mobius}.

We demonstrate how our boundary treatment performs on a square heightmap in Figure~\ref{fig:SGP2023}. Although previous works have already studied this application, we wanted to include a heightmap in our paper in case the reader is unfamiliar with these results. However, we note that --- to the best of our knowlege --- we are the first to propose this specific boundary treatment for heightmaps as well.

\section{Conclusion}

In this paper, we have shown how to generalize the moving mesh method to curved surfaces for use in shape approximation. We demonstrate that the shape operator can be used as a monitor function, and a projection term can account for curvature in a domain. Unlike previous surface meshing methods, this method keeps the mesh topology constant across iterations. 

\begin{wrapfigure}{r}{0.6\linewidth}
    \centering
    \vspace{-0.6\intextsep}
    \hspace*{-.6\columnsep}
    \includegraphics[width=\linewidth]{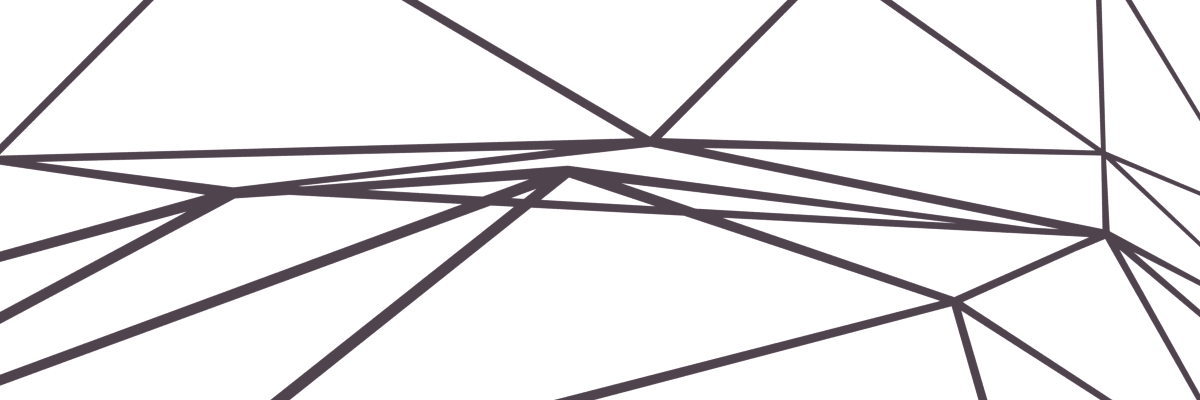}
    \caption{
    \label{fig:Flipped}%
        Example of a flipped element
    }
    \vspace{-\intextsep}
\end{wrapfigure}

Our 2D implicit Euler solver has a few limitations: first, the flow can become unstable and triangles can flip if triangles become too thin (see Figure~\ref{fig:Flipped}). This can happen if the surface has sharp corners and if \(\alpha\) is too high. 
We point out that the fandisks in Figures \ref{fig:Aniso2D} and \ref{fig:Aniso1D} needed to be smoothed out a bit in order for \(\cE_{2D}\) at to converge at high anisotropies.
The 1D spring graph solver is much more robust, however, it can still have difficulty converging if the mesh has corners that are too sharp. It also produces meshes that exhibit less anisotropy than those produced by the 2D energy.

As discussed in Section~\ref{sec:MeshAdaptationIn2D}, meshes that minimize (\ref{eq:DirichletEnergy}) aren't guaranteed to be equilateral under \(\sigma\); there may be some leftover anisotropy that accounts for stretch in the overall domain.

These problems pose interesting directions for future work. The fact that our flow doesn't require topology changes also makes it appealing for other applications. The anisotropic Laplacians we introduce could potentially be used in place of the typical graph or cotan Laplacians whenever adaptivity is required.

\bibliographystyle{eg-alpha-doi} 
\bibliography{bibliography}       



\newpage
\end{document}